\documentclass[seceq]{ptptex}

\usepackage{graphicx,color}

\title{
An Exotic Approach to Hadron Physics%
}

\author{
Masako \textsc{Bando}$^{1,}$\footnote{ e-mail address:bando@aichi-u.ac.jp
}
Yoko \textsc{Fukase}$^{2,}$\footnote{ e-mail address:g0540523@edu.cc.ocha.ac.jp
}
Jonathan \textsc{Shock}$^{3,}$\footnote{ e-mail address:jps@itp.ac.cn
}
Akio \textsc{Sugamoto}$^{2,}$\footnote{ e-mail address:sugamoto@phys.ocha.ac.jp
}
and
Sachiko \textsc{Terunuma}$^{2,}$\footnote{ e-mail address:g0570608@edu.cc.ocha.ac.jp
}
}

\inst{
$^1$ Physics Division, Aichi University, Aichi 470-0296, Japan\\
$^{2}$ Department of Physics, Ochanomizu Univeristy, Tokyo 112-8610, Japan\\
$^{3}$ Institute of Theoretical Physics, Beijing 100080, China
}

\abst{
An exotic approach to hadrons is discussed. It is based on the
recently developed open-closed string duality explicitly conjectured
as the AdS/CFT correspondence. Mesons as well as
pentaquarks are studied in this approach. Spins are introduced as
distribution functions over the string, and a second quantization
method of string theory is examined and used to estimate the mass
and decay width of various hadrons. This approach provides a way to
understand the structure of flavor by a configuration of probe
branes.}

\begin{document}

\maketitle

\section{Introduction}
The standard way to study hadrons is to estimate a hadron mass as
the sum of quark masses and spin-spin or hyperfine interactions.  To study the excitation states one uses a phenomenological linear plus Coulomb potential.

An exotic approach for this is to consider string theory in a curved
space which is the s-channel and t-channel dual to QCD\cite{M}, where spins
are distributed over the string.

Strings come in both open and closed form. Open strings have two end
points, each of which can be labelled with a Chan-Paton factor
giving a flavor or color quantum number. If both ends carry color,
the string transforms in the adjoint of the colour group, like the
gluon. If one end carries color and the other end flavor, the string
transforms in the fundamental of the color group and fundamental of
the flavour group, like a quark. If both ends carry flavor, the
string transforms as a meson. The classical energy of a string is
proportional to its length, so that, classically, the length of
massless gluons is vanishing, but those of heavier quarks and
mesons, corresponding to stretched strings, are not. The splitting
and merging of zero length strings at the color end gives the
emission and absorption of gluons, namely, the gauge interactions of
color (QCD). Similarly the splitting and merging of non-zero length
strings at the flavor end gives the gauge interaction of flavor,
where vector mesons, such as $\rho$ and K*, are gauge fields, but
the gauge symmetry is broken by their non-zero masses.

The graviton is one of the massless modes of the closed string. The
reason is as follows: an open string can make a small circle at any
point along the string, this circle can be split from the string,
propagates and is absorbed to another string. The probability of
emitting or absorbing of this circle is proportional to the length
(area) of the emitter or the absorber. Therefore, the closed string
coupling to an open string or an extended object is proportional to
the mass (length, area or volume) of the object. The closed string
has a number of modes, such as the graviton and the antisymmetric
field, but their couplings (or charge) to an extended object are
thus proportional to the mass of the object.

By placing a stack of $N_c$ color branes and $N_f (\ll N_c)$ flavor
branes in ten-dimensional space the respective Chan-Paton factors of
the strings attached to these branes give the correct transformation
properties for gluons, quarks and mesons. When the branes are
separated the stretched strings acquire a mass. By keeping the color
branes parallel the gluons remain massless but massive quarks and
hadrons can be introduced into this picture by the separation of the
flavor branes.

If the brane is the object of tieing the endpoint of open string.
The brane can be a textile made of the strings as threads.  Then,
the tension $T_p$ of a p-brane (the energy per unit volume of a
p-dimensionally extended object) is proportional to
$(T_1)^{(p+1)/2}$  by dimensional counting, where $T_1$ is the
tension of the original string.  The existence of massive color and
flavor branes distorts the extra space gravitationally. For
$N_f<<N_c$ we can treat this distortion as coming only from the
color branes and treat the flavor branes as probes. It is found that
the appropriate solution to the supergravity equations gives a space
which retains the Minkowski directions parallel to the D3 color
branes and warps the space in the directions perpendicular to these
branes.

Let us draw a number of circles on the color branes and flavor
branes, and draw a surface with these circles as its boundary. This
is a configuration of string theory contributing to a process of
QCD.  See for example Fig. 3 of Ref. 3).
\begin{figure}
\begin{center}
\includegraphics[width=10cm,clip]{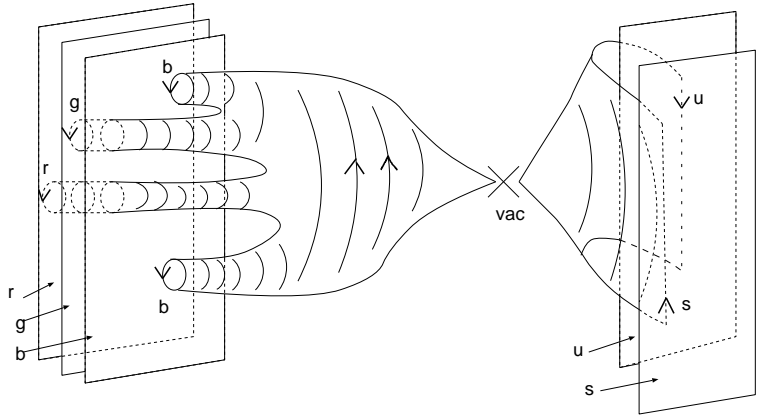}
\end{center}
\caption{Estimation of force between $u$ and $\bar s$.}
\label{duality}
\end{figure}
The s-channel view is to see the picture in the time direction
inside the brane. The t-channel view is to see this in the direction
of the extra space.  Here, we can understand that ``the number of
loops in the s-channel@view" (in the open string theory) is
replaced by ``the number of external lines in the t-channel view" (in
the closed string theory). Therefore, ``sum over loops in the open
string theory" is replaced by ``sum over external lines at tree level
of closed string theory", so that the non-perturbative QCD is
replaced by the classical gravity, where the sum over external lines
terminating on the color branes gives the classical contribution of
gravity from infinite points on the branes.  Of course the open
string theory and the closed string theory are different from QCD
and a certain gravitational theory, respectively. In order to obtain
classical supergravity decoupled from the flat-space region, we look
at the near horizon geometry with the string tension taken to
infinity, $N_c$ taken to infinity and the 't Hooft coupling $N_c
g_s$ taken to be a large constant. This classical theory of
supergravity corresponds to a strongly coupled large $N_c$ field
theory.\cite{M} To obtain QCD we need to keep $N_c=3$ and therefore to
consider the $1/N_c$ corrections properly.  Another important issue
is to break supersymmety by either turning on appropriate
supergravity fields or by embedding the branes in a configuration
which eliminates the supercharges.

A candidate for the t-channel view of realistic QCD is described in
a deformed space of ``6 dimensional AdS Schwarzshild space" x $S^4$.\cite{Kru}
In this model, color branes are spatially extended along the four
directions $({\bf x}_{\perp}, z, \vartheta)$, while flavor branes
are spatially extended along the six directions $({\bf x}_{\perp},
z, x^5, x^6, x^7)$. The $\vartheta$ direction is compactified on a
circle with radius $1/M_{KK}$ (a free parameter). The antiperiodic
boundary conditions of the fermions around the circle give
them a mass and break supersymmetry. The deformation of the space
comes only from the stuck $N_c$ heavy color branes, but not from
flavor branes. This is a reasonable assumption, since we want to
include the non-perturbative (multi-loop) effects of QCD, but not of
QFD.  (If we consider the gravitational@effects from the flavor branes, the non-perturbative effects of QFD would be included.)   

The metric of this deformed space is given by using $u=\sqrt{(x^5)^2+\cdots (x^9)^2}$,
\begin{equation}
ds^2=f(u)(-dt^2+dz^2+d{\bf x}_{\perp}^2)+g(u)du^2+g(u)^{-1}d\vartheta^2+f(u)^{-1}u^2 d\Omega^2_4,
\end{equation}
where
\begin{eqnarray}
f(u)=(u/R')^{3/2},~~g(u)=(f(u)h(u))^{-1},~~
h(u)=1-(U_{KK}/u)^3,  \label{QCD like model}
\end{eqnarray}
and $d\Omega^2_4$ is the metric on the $S^4$ with unit radius. $R'$
and $U_{KK}$ are given by using QCD coupling $\alpha_{c}$ as
\begin{equation}
R'^{3}=2\pi\alpha_{c}N_{c}/M_{KK},~~\mbox{and}~~ U_{KK}=\frac{8\pi}{9}\alpha_{c}N_{c}M_{KK}.
\end{equation}

\section{Pentaquark $\Theta^{+}$}
In this model, pentaquark baryons are studied.\cite{BKST} The
problem is to find catenary lines in the above deformed space, by
pinning down five end points of lines (strings) to the five flavors
$(udud\bar{s})$ on the respective flavor branes.  Here, the strings
should be connected at three junctions ($N_c=3$).  The mass formulae
was obtained roughly.  A bonus in this study is obtaining an
indication for the small decay width: In order for a pentaquark to
decay into a baryon and a meson, it should pass through a heavier
state with baryon or meson with a string loop. See the following Fig. 6 of Ref. 3).
\begin{figure}
\begin{center}
\includegraphics[width=12cm,clip]{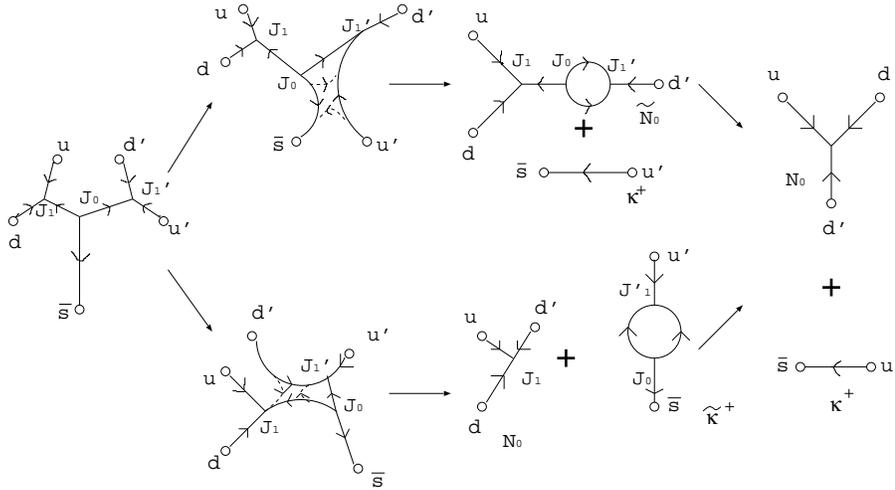}
\end{center}
\caption{Decay process of the pentaquark $\Theta^{+}$ in string theory.}
\label{decay}
\end{figure}
String theoretical estimation of the decay width following this figure is an open problem.

\section{Meson Strings and Flavor Branes}
Next, meson strings are studied in detail \cite{BSTmeson}.  There is a
clear difference between mesons with the same (mass) flavors $(u
\bar{u}, u \bar{d}, s \bar{s}, ...)$ and mesons with different
flavors $(u \bar{s}, u \bar{b}, d \bar{t}, ...)$.  The following
figures show that for mesons with different flavors, there is a
critical distance where the shape of the string connecting the
branes changes.
\begin{figure}
\begin{center}
\includegraphics[width=10cm,clip]{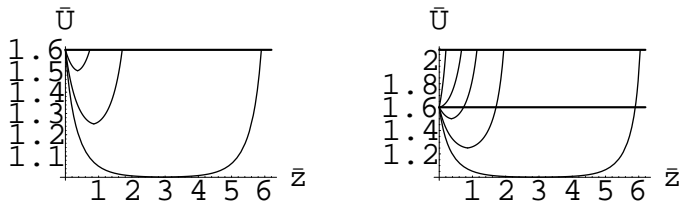}
\end{center}
\caption{Profile of $u\bar d$ (left) and $u\bar s$ (right) strings.} \label{udusstring}
\end{figure}
This critical point is also the point of inflection of the quark and
anti-quark potential.  
\begin{figure}
\begin{center}
\includegraphics[width=8cm,clip]{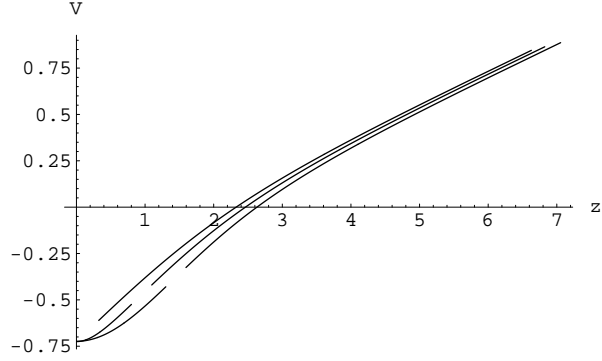}
\end{center}
\caption{Potential as a function of $r=Z\  {\rm GeV}^{-1}$ for the mesons  
$u\bar d$(the upper curve), $u\bar s$ (the middle curve) and $u\bar b$ (the lower curve).}
\label{potential}
\end{figure}
Namely, the potentials are different between
mesons with light flavors and those with heavy and light flavors.
Physical implications of this phenomenon are to be studied in
detail.

\section{Introduction of Spins into Hadrons}
We know from deep inelastic scattering experiment that quarks carry
only half of nucleon momentum and only 38 percent of nucleon spin.
The former is a common sense and the latter is a ``spin crisis"
\cite{Mak}.  Using string theory we may be able to find a solution
to this crisis, since both momentum and spin can be distributed over
strings (or hadron) by two variables, $X^{\mu}(\tau, \sigma)$ and
$\psi^{\mu}(\tau, \sigma)$. The $\psi$ satisfies a commutation
relation,
\begin{equation}
\{ \psi^{\mu}(\tau, \sigma), \psi^{\nu}(\tau, \sigma') \}=\eta^{\mu\nu}\delta(\sigma-\sigma'),
\end{equation}
so that $\psi^{\mu}(\tau, \sigma)/ \sqrt{2}$ can be a distribution
function of  $\gamma^{\mu}$ matrix. Therefore, spin is distributed
over the string theory.  The action in our deformed space may be
given, generalizing the work by Iwasaki and Kikkawa\cite{IK}, as
\begin{equation}
S=\int d^2\xi \sqrt{-det g_{ab}}~,
\end{equation}
with a world sheet metric
\begin{eqnarray}
g_{ab}=\partial_{a}X^{\mu}\partial_{b}X^{\nu}G_{\mu\nu}(x) +\frac{i}{2}\left(\bar{\psi}^{\mu}\hat{e}_{a}D_{b}\psi^{\nu}-D_{b}\bar{\psi}^{\mu}\hat{e}_{a}\psi^{\nu}\right)G_{\mu\nu}(x)~,
\end{eqnarray}
where $\hat{e}_{a}=e_{a\bar{a}}\gamma^{\bar{a}}$ is the contraction
of a zwiebein by a two dimensional $\gamma$ matrix, and
\begin{equation}
D_{a}\psi^{\mu}=\partial_{a}\psi^{\mu}+\partial_{a}X^{\nu}\Gamma^{\mu}_{\nu\lambda}(x)\psi^{\lambda}~.
\end{equation}
The action includes quadratic as well as quartic terms for
$\psi^{\mu}$, leading to the spin-spin or hyperfine interactions.  A
preliminary study\cite{BSTspin} shows that the energy of the meson
changes considerably by the excitation of spin degrees of freedom.

\section{Utilization of Second Quantization of Strings\cite{BFSS}}
To quantize a particle, we start from a constraint for its energy and momentum,
\begin{equation}
E-\frac{1}{2m}p^2-V(x)=0~,
\end{equation}
replace E by $i\frac{d}{dt}$ and p by $-i\frac{d}{dx}$ to satisfy
the commutation relations and apply the obtained constraint operator
to a wave function $\psi(t, x)$ of the particle. Then we arrive at
the quantum theory
\begin{equation}
\left[i\frac{d}{dt}-\frac{1}{2m}\left(i\frac{d}{dt}\right)^2-V(x)\right]\psi(t, x)=0~.
\end{equation}

In the same manner, (second) quantization of a string can be
performed with the constraint equations,
\begin{eqnarray}
P^{M} P_{M} + \Bigg( \frac{ 1 }{ 2 \pi \alpha ' } \Bigg)^{2} ( X^{ ' } )^{ 2 } = 0~,~~X^{'M} P_{M} = 0~.
\end{eqnarray}
The string is an extended object, so its momentum is a functional
differential, $P_{ M } ( \tau , \sigma ) = \frac{ - i \delta }{
\delta X^{ M } ( \tau , \sigma ) }$.  Now the wave equations of the
string $\Psi ( X^{ N } ( \tau , \sigma ) )$ can be obtained as
follows:
\begin{eqnarray} \label{wave equations}
\Bigg[ G^{ M N } ( x ) \Bigg( \frac{ - i \delta }{ \delta X^{ M } ( \tau , \sigma ) } \Bigg) \Bigg( \frac{ - i \delta }{ \delta X^{ N } ( \tau , \sigma ) } \Bigg) + \Bigg( \frac{ 1 }{ 2 \pi \alpha ' } \Bigg)^{2} G_{ M N } ( x ) ( X^{ ' M } ) ( X^{ ' N } ) \Bigg] \nonumber \\
\times \Psi ( X^{ N } ( \tau , \sigma ) ) = 0~,
\end{eqnarray}
\begin{eqnarray}
X^{ ' }_{ M } ( \tau , \sigma ) \Bigg( \frac{ - i \delta }{ \delta X^{ N } ( \tau , \sigma ) } \Bigg) \Psi ( X^{ M } ( \tau , \sigma ) ) = 0~.
\end{eqnarray}
Here the background metric $G_{ M N } ( x )$ is to be the deformed
one corresponding to the supergravity dual of large $N_c$ QCD.  We
know the classical static configuration of hadrons, such as mesons
as well as pentaquarks, in the deformed space as catenary lines.  As
we know, the wave function of the WKB approximation for a particle
takes the following form
\begin{equation}
\psi=e^{-i(Et+W_0+iW_1+W_2+iW_3+\cdots)},
\end{equation}
where $W_0$ is a classical action at rest, and the exponents of wave
function, being expanded in terms of $\hbar$, are alternately real
and imaginary. Therefore, in the string theory we set
\begin{eqnarray}
\Psi ( \tau , X^{I} ( \sigma ) ) = e^{-i \left(E\tau -P_{ z } z-\textbf{P}_{\perp} \cdot \textbf{x}_{ \perp } +E_{ cl } ( z )\tau \right) } \tilde{ \Psi } ( \tau , X^{I} ( \sigma ) ) ~,
\end{eqnarray}
 with
 \begin{equation}
 \tilde{\Psi}  ( \tau , X^{I} ( \sigma ) ) = e^{A-iB}~.
 \end{equation}
Here, $E_{ cl } ( z )$ is the static energy of a given string
configuration (depending on the perpendicular length $z$ of each
string segment), so that $E_{ cl } ( z )\tau$ is the classical
action of the string at rest, which corresponds to $W_0$ for a
particle. Therefore, the wave equations determine the real and
imaginary exponents A and B of wave function as quantum corrections.
So, we hope to determine the mass of hadrons as well@as the decay
width of them in this formulation. The existence of a critical point
found for a certain configurations of heavy and light mesons
suggests an unstable configuration at a saddle point, from which we
may estimate the decay rate of the meson as a tunneling probability.
We will clarify these points in the near future.

\section{Deformation of Flavor Branes (Running Behavior of Quark Mass)}
Under the gravitational force from the stuck $N_c$ color branes, the
flavor branes may change their shape.  This is a very important
issue.  The extra dimension $\lambda=\sqrt{(x^5)^2+\cdots+(x^7)^2}$
are internal coordinates of the flavor branes, while
$r=\sqrt{(x^8)^2+(x^9)^2}$ are perpendicular to the flavor brane
worldvolume. Before supersymmetries are broken, there is no force
between color branes and flavor branes and $r(\lambda)$ is constant
dynamically, but after the symmetries are broken flavor branes are
deformed in the UV as\cite{Kru}
\begin{eqnarray}
r(\lambda) \mathop{\longrightarrow}_{\lambda \to \infty}m_{\rm current}+\frac{\langle {\bar \psi}\psi\rangle}{\lambda}~.
\end{eqnarray}
Here, $r$ is the distance between color branes and the flavor brane,
so that it is related to the quark mass with this flavor. The value
of the quark bilinear condensate is zero in the supersymmetric case
but can be non-zero in the non-supersymmetric deformations. The
change of $r$ according to the change of $\lambda$ from $\infty$ to
zero is understood as a running behavior of the quark mass from UV
to IR. 
\begin{figure}
\begin{center}
\includegraphics[width=10cm,clip]{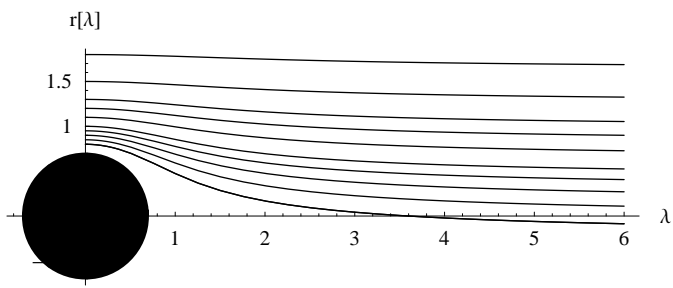}
\end{center}
\caption{Deformation of a D6-brane probe in the type IIA supergravity background
created by a stack of $N_c$ D4-branes with a compactified spatial dimension.}
\label{flows}
\end{figure}
By finding the well-behaved solutions for a given limiting value
in the UV, the quark bilinear condensate and quark mass can be read from
the UV asymptotics $r=m+\frac{c}{\lambda}$. The lowest line in the figure is a metastable flow which can be continuously deformed into a negative condensate solution. The renormalisation group flow of the quark mass is
still somewhat ill-defined in this holographic context though in Ref. 9) an attempt was made to define a gauge invariant
measure of the RG energy scale. The circular disk gives the region which
is not truly a part of the space due to the compactification.

\section{Energy of String Junction}
Another important issue is the energy of string junctions. In our study of
pentaquarks we have considered no energy for them. The energy of
junction in the QCD like model is estimated by Imamura\cite{Ima} as
\begin{equation}
E_J=c \frac{\pi N_c}{(2\pi)^4}u,~~c=-0.874~.
\end{equation}
This is -12 MeV, for $M_{KK}=0.77$ GeV which reproduces the Regge
slope of hadrons. Therefore, junctions may stabilize the exotic
fullerene or ferrule like compounds made of strings and junctions\cite{B}.

\section{Summary}
(1) An open string with color and flavor at its ends is a good
picture of a quark.

(2) Flavor structure may be understood in the picture of locating flavor branes in the extra space.

(3) Then, the shape of the potential between heavy and light quarks
gives a point of inflection.

(4) Spin may be distributed over the whole string.

(5) Treatment of kinetic terms in the estimation of hadron masses is
improved by the second quantization method of the string. Then, the
mass as well as the decay width of hadrons may be estimated.

(6) The radial direction perpendicular to the D3-branes is dual to
an energy scale in the field theory and the holographic
renormalisation group flow of the quark mass must be understood.

(7) There is evidence that the junction has negative energy.

(8) It is interesting to study the recently observed various exotic
hadrons, or the further exotics such as fullerene and ferrule with
our exotic approach given here. This is the end of our talk.

\section*{Acknowledgements}
The authors thanks Professor Kunihiro, Professor Oka, Professor Suganuma and the other members of the organizing committee of YKIS 2006 on "New Frontiers on QCD" for giving us an opportunity to present a talk and for giving us the good hospitality.




\begin{thebibliography}{99}

\bibitem{M} J. Maldacena, Adv. Theor. Math. Phys. {\bf 2} (1998), 231 [hep-th/9711200].

\bibitem{Kru} M. Kruczenski {\it et al.}, \JHEP{05,2004,41} [hep-th/0311270]. 

\bibitem{BKST} M. Bando, T. Kugo, A. Sugamoto and S. Terunuma, \PTP{112,2004,325}. 

\bibitem{BSTmeson} M. Bando, A. Sugamoto and S. Terunuma, \PTP{115,2006,1111}.

\bibitem{Mak} N. C. R. Makins, planary talk at ICHEP02, held at Amsterdam (2002).

\bibitem{IK} Y. Iwasaki and K. Kikkawa, \PRD {8,1973,440}.

\bibitem{BSTspin} M. Bando, A. Sugamoto and S. Terunuma, Poster presentation by Terunuma at Joint Meeting of Pacific Region Physics communities held at Hawaii (October 2006);\\
A. Sugamoto, hep-th/0502241.

\bibitem{BFSS} M. Bando, Y. Fukase, J. Shock and A. Sugamoto, Poster presentation by Fukase at Joint Meeting of Pacific Region Physics communities held at Hawaii (October 2006).

\bibitem{E} N. Evans, T. R. Morris and O. J. Rosten, \PLB{635,2006,148} [hep-th/0601114].

\bibitem{Ima} Y. Imamura, \PTP{115,2006,815}.

\bibitem{B} M. Bando, Talk at Miraikan in Tokyo (November 2004).


\end{thebibliography}
\end{document}